\documentclass[twocolumn,fp]{jpsj3}
\usepackage{graphicx}
\usepackage{txfonts,bm,revsymb}

\newcommand{\cop}{\hat{c}}
\newcommand{\Hop}{\hat{H}}

\newcommand{\veck}{\bm{k}}

\newcommand{\calH}{\mathcal{H}}
\newcommand{\calL}{\mathcal{L}}
\newcommand{\calW}{\mathcal{W}}

\begin{document}

\title{Impurity effects in a two-dimensional topological superconductor:
A link of $T_{\rm c}$-robustness with a topological number}

\author{Yuki Nagai, Yukihiro Ota, and Masahiko Machida}%
\inst{%
CCSE, Japan Atomic Energy Agency, 5-1-5 Kashiwanoha, Kashiwa, Chiba
 277-8587, Japan}%

\abst{
Impurity effects are probes for revealing an unconventional property in 
 superconductivity. 
We study effects of non-magnetic impurities, in
a 2D topological superconductor with $s$-wave pairing, the Rashba
spin-orbit coupling, and the Zeeman term. 
Using a self-consistent $T$-matrix approach, we calculate a
 phenomenological formula for the
 Thouless-Kohmoto-Nightingale-Nijs (TKNN) invariant in interacting
 systems, as well as density of 
 states, with different magnetic fields. 
This quantity weakly depends on the magnetic field, when a spectral gap
 opens, whereas this changes drastically, when in-gap states occurs. 
Furthermore, in the latter case, we find that the Anderson's
 theorem (robustness of $s$-wave superconductivity against non-magnetic
 impurities) breaks down. 
We discuss the origin, from the viewpoints of both unconventional
superconductivity and the TKNN invariant.
}

\maketitle

\section{Introduction}
Topological materials, such as semiconductors with 
the quantum Hall effect~\cite{Prange:1990} and topological
insulators~\cite{Hasan;Kane:2010} attract a great deal of
attention in condensed matter physics. 
Their essential character is classified by topological
invariants.~\cite{Nakahara:2003} 
Among them, topological
superconductors~\cite{Hor;Cava:2010,Wray;Hasan:2010,Sasaki;Ando:2011,Zhang;Jin:2011,Das;Kadowaki:2011} 
are notable materials, and their feature is
studied by different theoretical
ways.~\cite{Fu;Berg:2009,Sato:2010,Sato;Fujimoto:2010,Nagai;Machida:2013}  
Their topologically-protected features allow us to implement
applications, such as quantum
engineering.~\cite{Alicea;Fisher:2011}

The impurity effect in superconductivity is a probe for classifying
types of superconducting states, as well as tunneling
spectroscopy.~\cite{Kashiwaya;Tanaka:2000}  
Impurities leads to phenomena never occurring in
clean superconductors.~\cite{Kopnin:2001}
A gapless behavior in the density of states (DOS) via non-magnetic
impurity scattering,~\cite{Hotta:1993,Preosti;Muzikar:1996} for example,
is witness of unconventional superconductivity. 

One of the celebrated statements about superconducting alloys is the
Anderson's theorem~\cite{Kopnin:2001};
the robustness of $T_{\rm c}$ in $s$-wave superconductivity against
non-magnetic impurities. 
This prediction is related to the absence of low-energy excitations in
\textit{conventional} superconductivity. 
Here, we pose a simple question: Is a topological superconductor
conventional? 
Intuitively, such a superconductor should be robust against
with impurities, from its topological nature. 
Moreover, if the superconductivity occurs under $s$-wave pairing, the
Anderson's theorem implies that this material is free from non-magnetic
impurities. 
A topological $s$-wave superconductor is predicted by Sato, Takahashi,
and Fujimoto.~\cite{Sato;Fujimoto:2010, SatoPRL} 
We notice that, however, this prediction indicates the system is 
regarded as a chiral $p$-wave model; 
the topological $s$-wave superconductivity can have
\textit{unconventional} features. 
To construct theory of dirty topological superconductors is
desirable for answering this intriguing issue. 
This approach is also useful for development of quantum
engineering based on topological superconductors, since typical
materials in experiments would be dirty. 

\begin{figure}[bh]
\begin{center}
     \begin{tabular}{p{ 0.5 \columnwidth}} 
      \resizebox{0.5 \columnwidth}{!}{\includegraphics{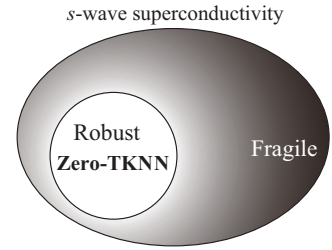}} 
    \end{tabular}
\end{center}
\caption{\label{fig:circle} Schematic diagram of our main
 result; robustness of $T_{\rm c}$ against non-magnetic impurities, in
 an $s$-wave superconductor with the Rashba spin-orbit
 coupling and the Zeeman magnetic field.~\cite{Sato;Fujimoto:2010}  
The brighter region (zero-TKNN) is more robust than the darker
 region. 
Inside the zero-TKNN region, a topological invariant,
 Thouless-Kohmoto-Nightingale-Nijs (TKNN) invariant is zero, without
 impurities. 
Otherwise, the robustness reduces continuously. }
\end{figure}

In this paper, we study effects of non-magnetic impurities, in
a topological superconductor with $s$-wave pairing, the Rashba
spin-orbit coupling, and the Zeeman term.~\cite{Sato;Fujimoto:2010} 
Our idea for revealing the impurity effects is to use a formula relevant
to Thouless-Kohmoto-Nightingale-Nijs (TKNN)
invariant,~\cite{Thouless;Nijs:1982,Kohmoto:1985} in addition to 
standard arguments in superconducting alloys. 
We obtain this formula phenomenologically, using an expression for the
TKNN invariant derived by Niu, Thouless, and Wu.~\cite{Niu;Wu:1985}  
We call it NTW-TKNN formula. 
The value of this formula is a leading term of a generic TKNN
invariant in interacting
systems,~\cite{Ishikawa;Matsuyama:1987,Volovik:2003,Gurarie:2011} under
momentum-independent self-energy. 
We find that its numerical calculations are more stable than the generic
formula, when the superconducting pair breaking occurs via non-magnetic
impurities. 
The NTW-TKNN formula and low-energy behaviors in DOS (the presence of in-gap
states) are calculated by the self-consistent $T$-matrix approximation, 
changing the magnetic field. 
The critical temperature $T_{\rm c}$ is evaluated by self-consistent
calculations of the gap equation. 

The $T_{\rm c}$ reduction behavior in this topological superconductor 
is characterized well in terms of the NTW-TKNN formula
When a spectral gap opens in the DOS under impurities,
the NTW-TKNN formula quite weakly depends on the magnetic field. 
We find that a robust topological superconducting state emerges in this
magnetic-field region.  
In contrast, when in-gap states occur in the DOS, this quantity depends
on it strongly. 
In this case, we have a topological superconducting state fragile
against non-magnetic impurities. 
We also find a trivial state, whose NTW-TKNN formula in clean limit
(i.e., the TKNN invariant) has a non-zero even value, in the latter
magnetic-field region. 
This result indicates that an $s$-wave state is suffered
from non-magnetic impurities. 
Therefore, we find that robustness of $s$-wave superconductivity
against non-magnetic impurities (Anderson's theorem) breaks down, when
the TKNN invariant has a non-zero value (See, Fig.~\ref{fig:circle}). 
In other words, a class of $s$-wave superconductivity is
sensitive to non-magnetic impurities, like unconventional
superconductivity. 

The origin of this violation is clarified from two viewpoints. 
First, we find a similar mechanism to chiral $p$-wave. 
The fragile $T_{\rm c}$ is related to the self-energy from
impurity scattering. 
Second, the low-energy excitations related to gapless edge modes, whose
 number is characterized by the TKNN
 invariant,~\cite{Sato;Fujimoto:2010} is relevant to the
 reduction of $T_{\rm c}$. 
Thus, we find a link of $T_{\rm c}$ with a topological number, under
non-magnetic impurities. 

The paper is organized as follows. 
Section \ref{sec:model} shows the Hamiltonian for the system. 
Section \ref{sec:formulation} shows our approach to examine the
impurity effects. 
Section \ref{sec:results} is the main part of this paper. 
Calculating the NTW-TKNN formula and the DOS, we show a connection of a
topological number with the presence of in-gap states.  
We discuss the origin of the violation of the Anderson's theorem in
Sec.~\ref{sec:discussion}. 
Section \ref{sec:summary} is devoted to summary. 

\section{Model}
\label{sec:model}
The mean-field Hamiltonian~\cite{Sato;Fujimoto:2010} is 
\(
\Hop = (1/2)\sum_{\veck} 
\hat{\Psi}_{\veck}^{\dagger} 
\calH(\veck)
\hat{\Psi}_{\veck}
\), with \(
\hat{\Psi}_{\veck}
=
(
\cop_{\veck,\uparrow},\,
\cop_{\veck,\downarrow},\,
\cop_{-\veck,\uparrow}^{\dagger},\,
\cop_{-\veck,\downarrow}^{\dagger}
)^{\rm T}
\). 
The creation and the annihilation operators of an electron with momentum
$\bm{k}=(k_{x},k_{y})$ in the first Brillouin zone and spin
$\sigma\,(=\uparrow,\downarrow)$ are, respectively,  
\(
\cop^{\dagger}_{\veck,\sigma}
\) 
and 
\(
\cop_{\veck,\sigma}
\). 
The $4\times 4$ Bogoliubov-de Gennes (BdG) Hamiltonian is 
\begin{equation}
\calH(\veck)
=
\left(
\begin{array}{cc}
h_{0}(\veck) &  i \Delta \sigma_{2}  \\
-i \Delta^{\ast} \sigma_{2} & -h_{0}^{\ast}(-\veck)
\end{array}
\right), 
\end{equation}
with 
\(
h_{0}(\veck)
=
\varepsilon(\veck)
-
h \sigma_{3} + \alpha \calL(\veck)
\). 
The band dispersion is 
\(
\varepsilon(\veck) = -2t(\cos k_{x} + \cos k_{y}) - \mu
\), with hopping parameter $t$ and chemical potential $\mu$. 
The Zeeman magnetic field is $h$. 
The symbol $\sigma_{i}$ is the $i$th component of the $2\times 2$ Pauli
matrices ($i=1,\,2,\,3$). 
The Rashba spin-orbit coupling is described by 
\(
\alpha \calL
=
\alpha (\sigma_{1}\sin k_{y} - \sigma_{2}\sin k_{x})
\), with $\alpha >0$. 
Throughout this paper, we set $\alpha=t$. 
The off-diagonal block of $\calH$ means $s$-wave pairing interaction, 
\(
\sum_{\veck}
(
\Delta 
\cop^{\dagger}_{\veck,\uparrow}\cop^{\dagger}_{-\veck,\downarrow}
+
\text{h.c.}
) 
\). 
The BdG Hamiltonian and the eigenvalues are periodic 
with respect to $\veck$. 
However, the eigenvectors are not so, since the overall phases are 
not uniquely determined by the eigenvalue equation.  
Thus, we find $U(1)$
principle bundle on torus,~\cite{Kohmoto:1985} where the first
Brillouin zone is identified with torus. 

Using an analogy with the integer quantum Hall effect, the TKNN
invariant (i.e., the first Chern class~\cite{Nakahara:2003}) is
examined for the dichotomy of the superconducting
state.~\cite{Sato;Fujimoto:2010} 
The strength of the Zeeman magnetic field is crucial for
the emergence of the topological superconductivity. 
The topological order also relies on the characteristic band
structure of the normal state. 
The Rashba coupling shifts the band minimum of $\varepsilon(\veck)$, 
depending on chirality. 
These shifted bands have crossing points each other around the original
band minimum, and the Zeeman energy induces an energy gap there. 
In contrast to a pure Zeeman-field case, the spin states in the
resultant bands have both spin-up and spin-down components. 
This feature is similar to a topological order in a semiconductor on a $s$-wave
superconductor.~\cite{Alicea;Fisher:2011} 

\section{Formulation}
\label{sec:formulation}
\subsection{NTW-TKNN formula}
We show a formula related to the TKNN invariant, i.e.,
NTW-TKNN formula, to understand the impurity effects in a
topological superconductor. 
This is defined by
\begin{equation}
\calW
= 
\int \! \frac{d^{2}\veck}{2\pi} 
\int_{-\infty}^{\infty}\!\! \frac{d\omega}{2\pi}\,
 \tr\bigg[
i\,G_{\veck}(i\omega)
\frac{\partial \calH}{\partial k_{x}}
G_{\veck}(i\omega)
\frac{\partial \calH}{\partial k_{y}}
G_{\veck}(i\omega) 
\bigg], \label{eq:tkkn}
\end{equation}
with the Green's function 
\(
 G_{\veck}(\Omega) 
= 
[\Omega - \calH(\veck) -  \Sigma(\Omega)]^{-1}
\). 
The $\veck$-integral is taken over the first Brillouin zone. 
The effect of impurity scattering is taken as the self-energy $\Sigma$. 
In our model the self-energy does not depend on $\veck$, since the
impurities are randomly distributed. 
Although this $\veck$-independent model is a restricted way to treat
impurity scattering in superconductors, this approach is widely used for
revealing impurity effects in different dirty-limit
superconductors~\cite{Kopnin:2001,Hotta:1993,Preosti;Muzikar:1996} and 
is a reasonable starting point of discussing $T_{\rm c}$-robustness. 
We obtain formula (\ref{eq:tkkn}) phenomenologically, using an
expression of the TKNN invariant derived by Niu, Thouless, and
Wu;~\cite{Niu;Wu:1985} we replace single-particle Green's
functions in their expression with $G_{\veck}(\Omega)$.  
In contrast to the TKNN invariant, $\calW$ is a continuous real number. 
This formula is regarded as a leading term of a generic topological
invariant in interacting
systems.~\cite{Ishikawa;Matsuyama:1987,Volovik:2003,Gurarie:2011}
We will address the calculations of the generic formula in
Sec.~\ref{subsec:genericTKNN}.  

\subsection{Self-consistent $T$-matrix approach}
Our approach for evaluating the self-energy $\Sigma$ is
the self-consistent $T$-matrix approximation. 
The $T$-matrix for randomly distributed non-magnetic
impurities~\cite{Mahan:2000} is
\begin{equation}
T(\Omega) 
= \left[ 
1 - 
V 
\frac{1}{N}\sum_{\veck} G_{\veck}(\Omega)
\right]^{-1} V, 
\label{eq:tmatrix}
\end{equation}
with \mbox{$V = {\rm diag} \: (V_{0},V_{0},-V_{0},-V_{0})$} 
and 
the $\veck$-mesh size $N$. 
The self-energy~\cite{note:selfenergy} is
\begin{equation}
\Sigma(\Omega) 
= n_{\rm imp} T(\Omega)-n_{\rm imp}V, \label{eq:S}
\end{equation}
with impurity concentration $n_{\rm imp}$. 
The anomalous ($f_{\veck}$) and the normal ($g_{\veck}$) Green's
functions are $2\times 2$ block matrices of $G_{\veck}(\Omega)$, 
\begin{equation}
G_{\veck}(\Omega) = \left(\begin{array}{cc}
g_{\veck}(\Omega) & f_{\veck}(\Omega)  \\
f^{\dagger}_{\veck}(\Omega) & \bar{g}_{\veck}(\Omega)
\end{array}\right).
\label{eq:G}
\end{equation}
The pair potential $\Delta$ is evaluated by the gap
equation
\begin{equation}
i\Delta \sigma_{2} = V_{\rm int} \frac{T}{N} \sum_{n=-n_{\rm c}}^{n_{\rm c}} \sum_{\veck} f_{\veck}(i \omega_{n}), \label{eq:delta}
\end{equation}
with pairing interaction strength $V_{\rm int}$ and the fermionic Matsubara
frequency $\omega_{n} = \pi T (2 n + 1)$. 
We use the cutoff parameter $n_{\rm c}$ such that 
$\omega_{n_{\rm c}} =  10 \pi$. 
The DOS is 
\begin{equation}
N(E) 
= -\frac{1}{2\pi} \frac{1}{N}\sum_{\veck}\,\tr[ {\rm Im}\, g_{\veck}(E)].
\end{equation}
The evaluation of $T_{\rm c}$ depending on $n_{\rm imp}$ is 
performed, with fully self-consistent calculations, i.e., solving
Eqs.~(\ref{eq:tmatrix})--(\ref{eq:delta}) self-consistently. 
In the calculations of $\calW$ and $N(E)$, we self-consistently
solve~\cite{note:calculations} 
Eqs.~(\ref{eq:tmatrix})--(\ref{eq:G}) for given $\Delta$ and
$\mu$, to compare our results with the arguments by Sato et
al.\cite{Sato;Fujimoto:2010}. 

\section{Results}
\label{sec:results}
\subsection{NTW-TKNN formula and density of states}
\begin{figure}
\begin{center}
     \begin{tabular}{p{ \columnwidth}} 
      \resizebox{\columnwidth}{!}{\includegraphics{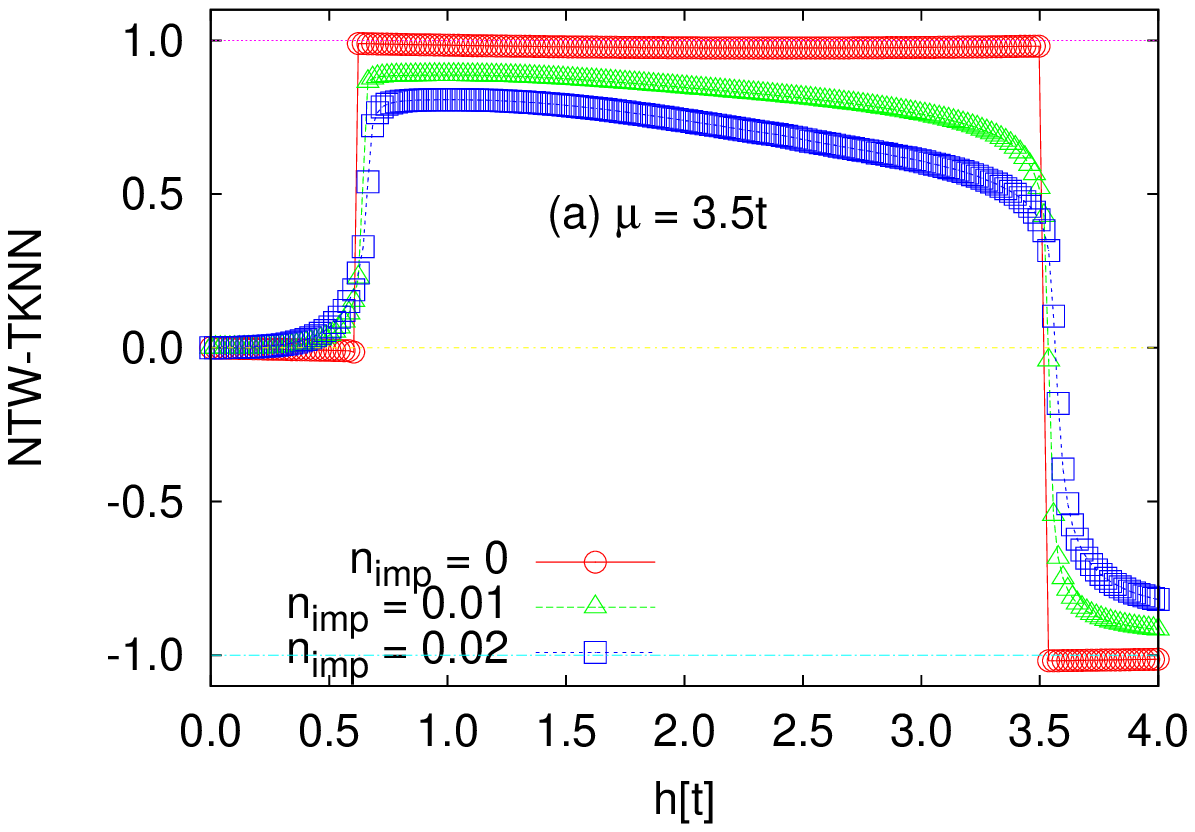}} \\ 
      \resizebox{\columnwidth}{!}{\includegraphics{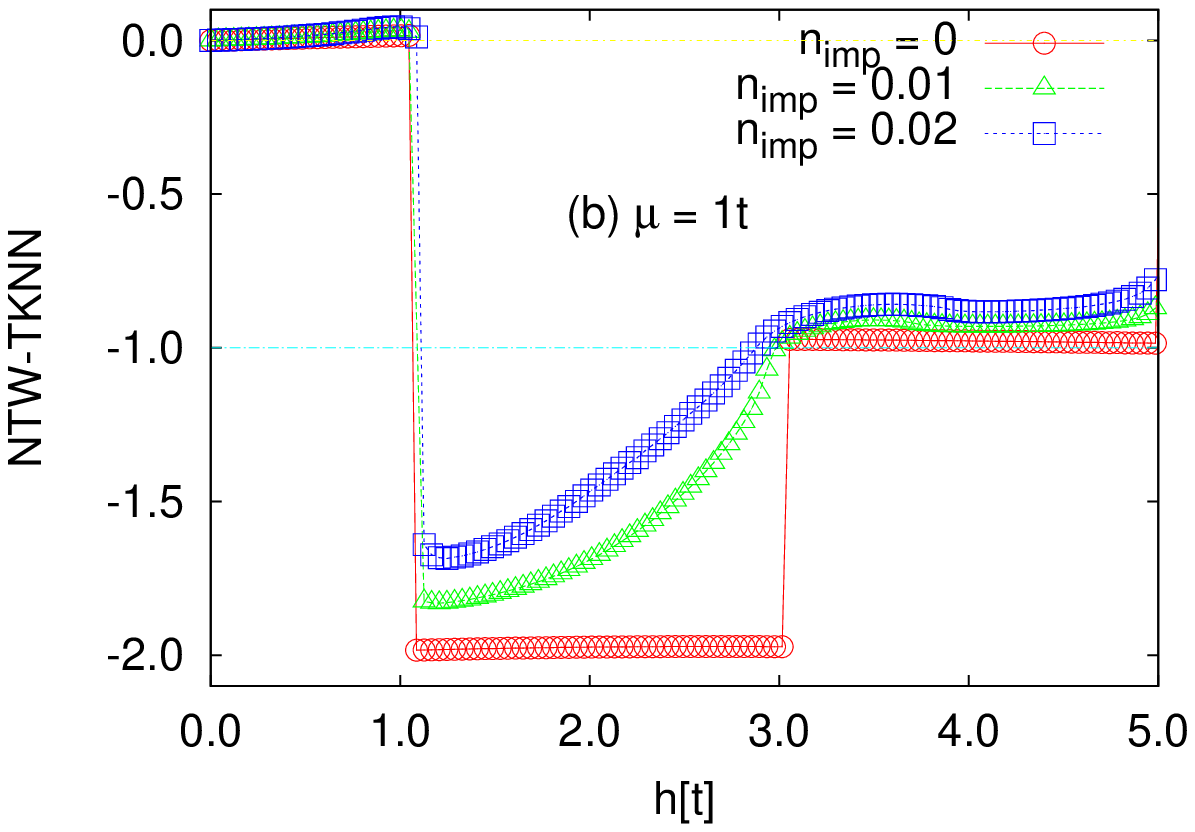}} 
    \end{tabular}
\end{center}
\caption{\label{fig:tkkn}(Color online) Zeeman-magnetic-field dependence
 of NTW-TKNN formula (\ref{eq:tkkn}), with different
 impurity concentrations ($n_{\rm imp}=0,\,0.01,\,0.02$). 
The horizontal axis is magnetic field $h/t$, with hopping parameter
 $t$. 
The chemical potentials are (a) $\mu = 3.5t$ and (b) $\mu = 1t$. 
The pair potential is $\Delta = 0.35t$. }
\end{figure}

Now, we show the results for the NTW-TKNN formula. 
Figure \ref{fig:tkkn} shows the magnetic-field dependence, with
different impurity concentrations. 
The impurity strength is $V_{0}=3t$ and the pair potential is 
$\Delta = 0.35 t$. 
We examine two kinds of the chemical potential, $\mu=3.5t$ and
$\mu=t$. 
In Eq.~(\ref{eq:tkkn}), the number of the meshes for the
$\veck$-summation is $480 \times 480$, and the energy-integration range
is $- 2.5t \le \omega \le 2.5t$.  
Without impurities ($n_{\rm imp} = 0$), the NTW-TKNN formula takes an 
integer for each magnetic field. 
This result is consistent with the exact analysis.~\cite{Sato;Fujimoto:2010} 
The TKNN invariant is an even number for a non-topological magnetic-field
region.~\cite{Sato;Fujimoto:2010} 
Otherwise, a topological superconducting state occurs. 
In the zero TKNN-invariant region, $\calW$ is
not significantly changed, compared to 
the case when $\calW|_{n_{\rm imp}=0}$ takes a non-zero value. 
We can find that the DOS is robust against the non-magnetic impurities
in this region, as well. 
These behaviors correspond to the Anderson's theorem. 

\begin{figure}
\begin{center}
     \begin{tabular}{p{\columnwidth}}
      \resizebox{\columnwidth}{!}{\includegraphics{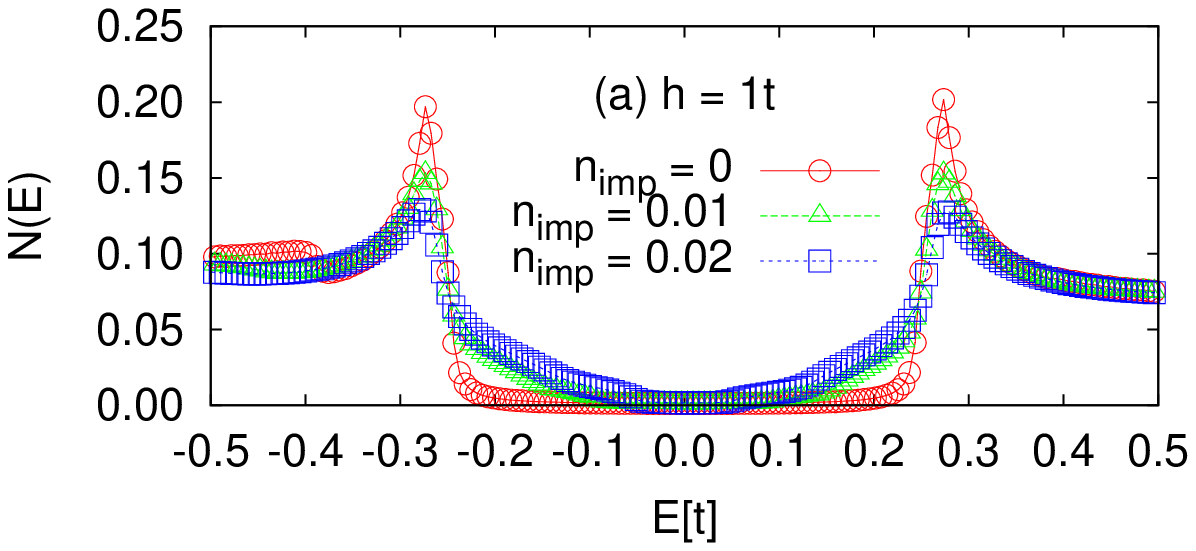}} \\
      \resizebox{\columnwidth}{!}{\includegraphics{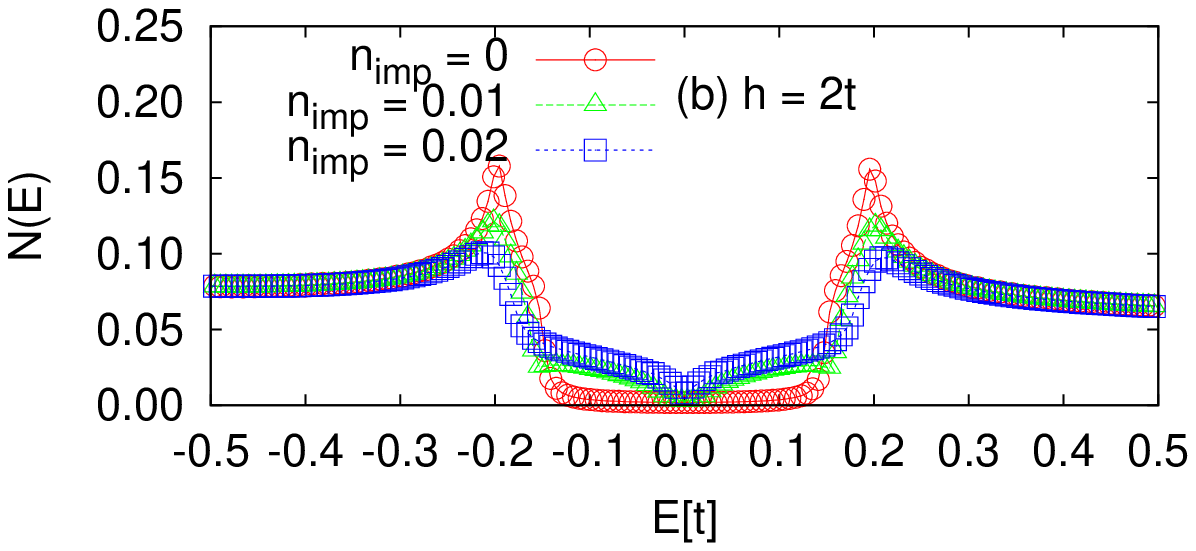}} 
    \end{tabular}
\end{center}
\caption{\label{fig:DOS}(Color online) Energy dependence of density
 of states $N(E)$, with different impurity concentrations 
($n_{\rm  imp}=0.0,\,0.01,\,0.02$).  
The horizontal axis is energy $E/t$, with hopping parameter $t$. 
The Zeeman magnetic fields are (a) $h = 1t$ and (b) $h = 2t$. 
 The other parameters are the same as in Fig.~\ref{fig:tkkn}(a). }
\end{figure}

\begin{figure}
\begin{center}
     \begin{tabular}{p{\columnwidth}}
      \resizebox{\columnwidth}{!}{\includegraphics{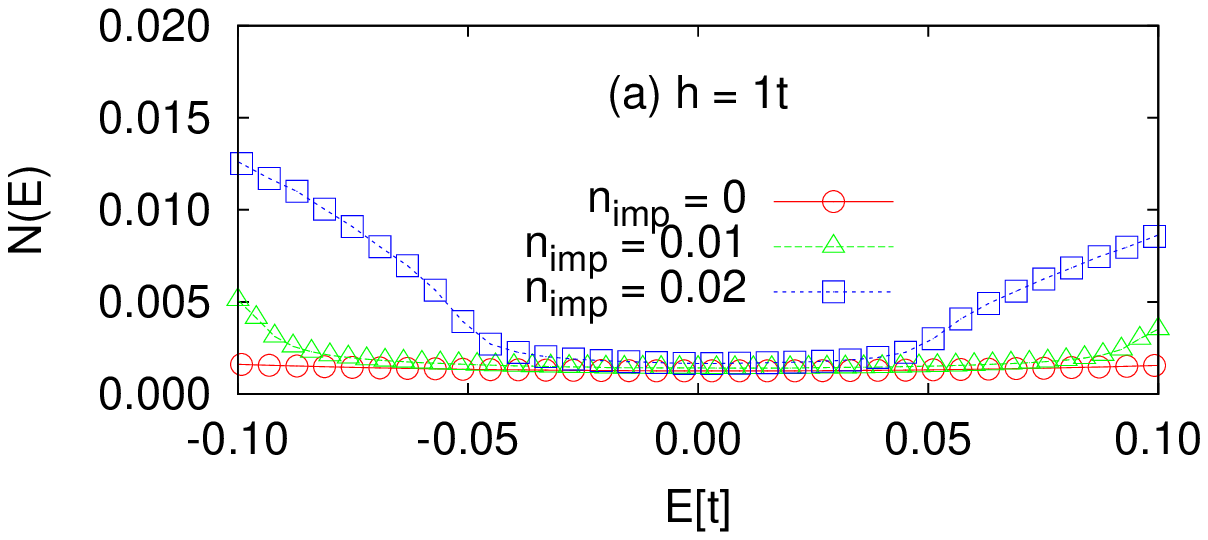}} \\
      \resizebox{\columnwidth}{!}{\includegraphics{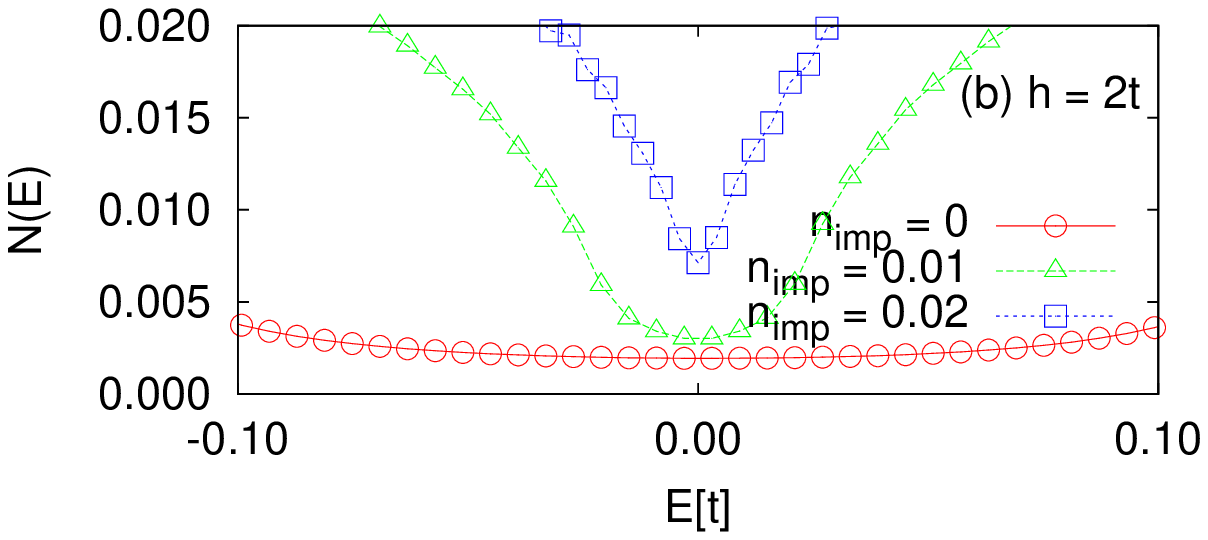}} 
    \end{tabular}
\end{center}
\caption{\label{fig:exDOS}
(Color online) Low-energy behaviors of density
 of states $N(E)$ (a) in Fig.~\ref{fig:DOS}(a) and (b)
 in Fig.~\ref{fig:DOS}(b), with different impurity concentrations 
($n_{\rm imp}=0.0,\,0.01,\,0.02$). 
The horizontal axis is energy $E/t$, with hopping parameter $t$. 
(a) The spectral gap definitely opens for every impurity concentration. 
(b) To compare with $n_{\rm imp}=0$, in-gap states occur for
 $n_{\rm imp} \neq 0$. }
\end{figure}

Let us focus on the $1$-TKNN-invariant region in Fig.~\ref{fig:tkkn}(a)
($0.5 t <h <3.5 t$). 
We find that the NTW-TKNN formula weakly depends on $h$ but is almost
constant up to $h \sim 1 t$, whereas $\calW$ strongly depends on
$h$ (a decreasing behavior) for $h > 2 t$.   
For understanding these behaviors more clearly, we turn into the results for
the DOS (Fig.~\ref{fig:DOS}), with $\mu = 3.5 t$. 
In the calculations of the DOS, the $\veck$-mesh size is 
$960\times 960$. 
Figures \ref{fig:DOS}(a) and \ref{fig:exDOS}(a) show that the spectral
gap opens for 
every impurity concentration, when the change of the NTW-TKNN formula is
moderate against $h$ (i.e., $h=1 t$). 
However, when $\calW$ is decreasing ($h > 2t$), the spectral gap closes
and in-gap states occurs for high impurity concentration ($n_{\rm
imp}=0.02$), as shown in Figs.~\ref{fig:DOS}(b) and \ref{fig:exDOS}(b).  


\subsection{Robustness of $T_{\rm c}$}
\begin{figure}
\begin{center}
     \begin{tabular}{p{\columnwidth}}
      \resizebox{\columnwidth}{!}{\includegraphics{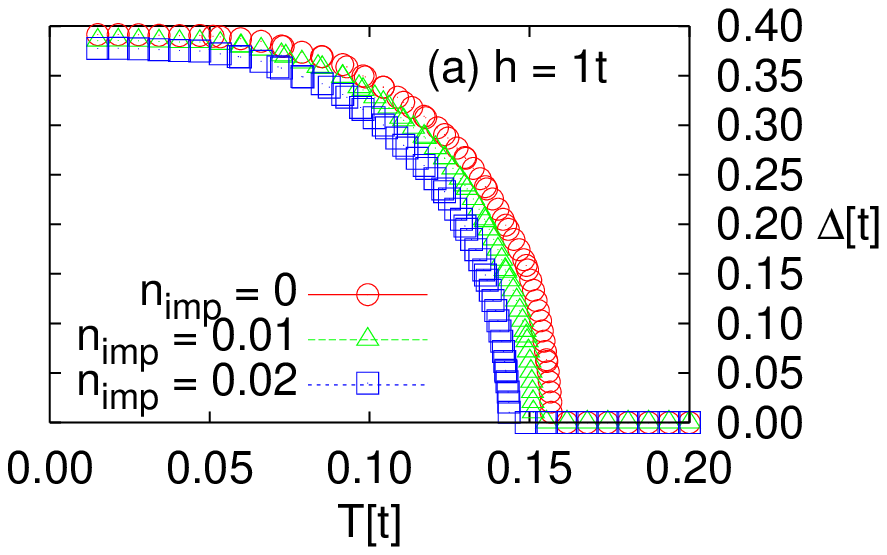}} \\
      \resizebox{\columnwidth}{!}{\includegraphics{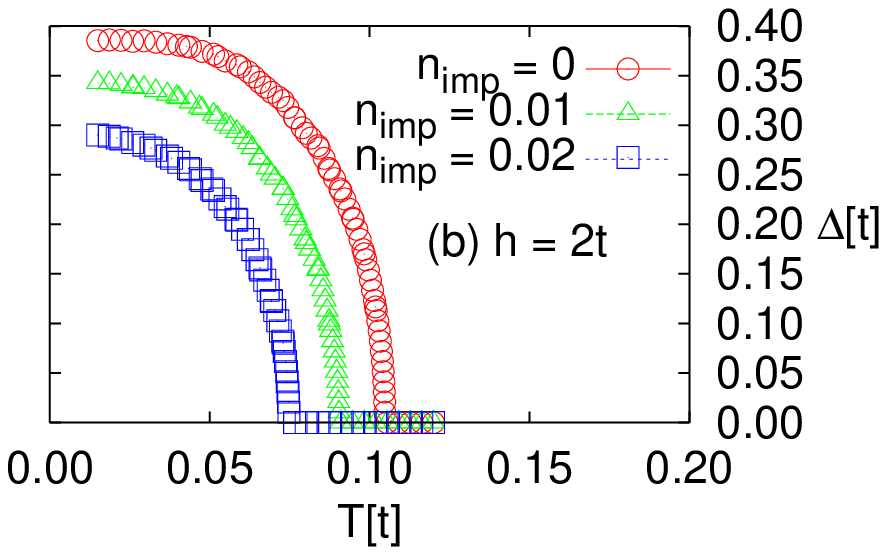}} 
    \end{tabular}
\end{center}
\caption{\label{fig:tc}(Color online) Temperature dependence of pair
 potential $\Delta$, with different impurity 
 concentrations ($n_{\rm imp}=0.0,\,0.01,\,0.02$). 
The chemical potential is $\mu=3.5\,t$, with hopping parameter $t$. 
The magnetic field $h$ and the pairing interaction strength 
$V_{\rm int}$ are (a) $(h,V_{\rm int})=(1t,\,5.6t)$ 
and (b) $(h,V_{\rm int})=(2t,\,8t)$. The critical temperature such that
 $\Delta =0$ for (a) is not dominated by $n_{\rm imp}$, compared to (b).}
\end{figure}

Now, we examine the robustness of $T_{\rm c}$ against non-magnetic
impurities, in terms of the NTW-TKNN formula. 
We can find that $T_{c}$ is robust for the zero-TKNN-invariant
regions [$0 < h <0.5t$ in Fig.~\ref{fig:tkkn}(a) and $0<h<1 t$ in
Fig.~\ref{fig:tkkn}(b)]. 
We confirm the Anderson's theorem, again. 
When the topological $s$-wave superconducting state (i.e., a state with
an odd TKNN invariant) occurs, the behaviors of
$T_{\rm c}$ against $n_{\rm imp}$ are strongly correlated with the
NTW-TKNN formula. 
Figure \ref{fig:tc}(a) shows that $T_{\rm c}$ is not so reduced, with
increasing the impurity concentrations, when $\calW$ is almost constant
[$h=1 t$ in Fig.~\ref{fig:tkkn}(a)]. 
However, when $h= 2t$ (i.e., the value of the NTW-TKNN formula differs
from a constant), $T_{\rm c}$ is fragile 
against $n_{\rm imp}$, as seen in Fig.~\ref{fig:tc}(b). 
This behavior is compatible with the
prediction~\cite{Sato;Fujimoto:2010} that a chiral $p$-wave-like state
occurs. 
Thus, in the magnetic-field region for an odd TKNN invariant, the
robustness of $T_{\rm c}$ against non-magnetic impurities reduces
gradually. 
Furthermore, we stress that a fragile behavior of $T_{\rm c}$ occurs
even in a non-topological state.  
We focus on the even TKNN-invariant region in Fig.~\ref{fig:tkkn}(b) 
($1 t< h < 3 t$). 
A non-topological superconducting state occurs in this region,
without non-magnetic impurities.~\cite{Sato;Fujimoto:2010}
Figure \ref{fig:tkkn}(b) shows that, however, the value of the NTW-TKNN
formula  differs from the constant value $-2$. 
We can find that $T_{\rm c}$ reduces, with increasing $n_{\rm imp}$. 
To sum up, the Anderson's theorem
for an $s$-wave superconductivity breaks down, when the NTW-TKNN formula has
a non-zero value. 

\subsection{Generic TKNN invariant for non-zero spectral gap}
\label{subsec:genericTKNN}
\begin{figure}
\begin{center}
     \begin{tabular}{p{ \columnwidth}} 
      \resizebox{\columnwidth}{!}{\includegraphics{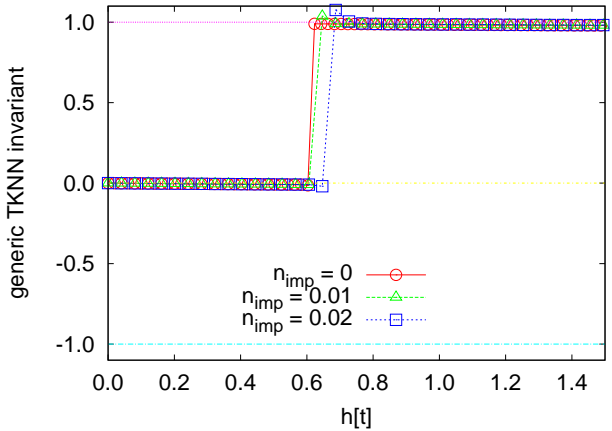}} \\ 
    \end{tabular}
\end{center}
\caption{\label{fig:gentknn}
(Color online) Generic TKNN invariant in a low magnetic-field region,
 with different impurity concentrations ($n_{\rm imp}=0,\,0.01,\,0.02$). 
The horizontal axis is
 magnetic field $h/t$, with hopping parameter $t$. The chemical
 potential is $\mu = 3.5t$. The pair potential is $\Delta = 0.35t$. 
This parameter set is the same as in Fig.~\ref{fig:tkkn}(a). }
\end{figure}

The previous results indicate that the
superconducting order survives well, when $h < 2\,t$. 
Let us consider this claim, in terms of the generic TKNN invariant 
in interacting
systems.~\cite{Ishikawa;Matsuyama:1987,Volovik:2003,Gurarie:2011} 
Under the condition that the self-energy $\Sigma$ is independent of
$\veck$, the relationship of the generic formula $\calW_{\rm gen}$ to $\calW$ 
is $\calW_{\rm gen}  = \calW + \delta \calW$, with 
\begin{eqnarray}
&&
\delta \calW  
=
\int \! \frac{d^{2}\veck }{2\pi} 
\int_{-\infty}^{\infty} \! \frac{d\omega}{2\pi}\,\,
 \tr \bigg[
(-1)\frac{d \Sigma(i\omega)}{d\omega}
\nonumber \\
&&
\quad\quad\quad\quad\quad
\times 
\,G_{\veck}(i\omega)
\frac{\partial \calH}{\partial k_{x}}
G_{\veck}(i\omega)
\frac{\partial \calH}{\partial k_{y}}
G_{\veck}(i\omega) 
\bigg]. 
\end{eqnarray} 
Figure \ref{fig:gentknn} shows the magnetic-field dependence of
$\calW_{\rm gen}$ for $\mu=3.5\,t$, $\Delta = 0.35\,t$, and 
$h < 2\,t$, with different impurity concentrations. 
The physical parameter set is the same as in Fig.~\ref{fig:tkkn}(a). 
We find that the topological feature without
non-magnetic impurities does not change, even in the presence of
impurity scattering. 
We can find that the calculations of $\calW_{\rm gen}$ are unstable
when the spectral gap is closed (e.g., $h > 2\,t$).  
This difficulty may come from a singular behavior of 
\(
d\Sigma(i\omega)/d\omega
\) on the imaginary axis. 
In addition, we speculate that the in-gap state
excitations shown in Fig.~\ref{fig:exDOS}(b) can be an obstacle 
to assess a topological property via $\calW_{\rm gen}$. 
Berry's connection $1$-form built up by the occupied bands (negative
eigenstates) of the BdG Hamiltonian would be ill-defined, since the
in-gap states can induce transition from the negative eigenstates
and to the positive ones, and vice versa.  

\section{Discussion}
\label{sec:discussion}
We argue the origin of the fragile behaviors of $T_{\rm c}$ and the
violation of the Anderson's theorem. 
First, we consider this point to be typical arguments of the impurity
effects in unconventional superconductors. 
The Anderson's theorem breaks down when the
$\veck$-averaged anomalous self-energy vanishes~\cite{Kopnin:2001}.  
Let us use the non-self-consistent Born approximation, for
simplicity. 
The anomalous self-energy is 
\(
\Sigma^{\rm A}_{\rm Born}(\Omega) 
= -( n_{\rm imp}V_{0}^{2} /N) \sum_{\veck} f_{\veck}(\Omega)
\).
This quantity vanishes in chiral $p$-wave superconductors. 
Now, we obtain
\begin{equation}
\Sigma^{\rm A}_{\rm Born}(\Omega)
=
i \Delta \sigma_{2} \frac{n_{\rm imp} V_{0}^{2} }{N}
\sum_{\veck}D^{-1}(\veck)
[C(\veck) - (h-\Omega)^{2}],
\end{equation}
with
\(
C = \alpha^{2}|\mathcal{L}_{12}|^{2} + |\Delta|^{2} + 
\varepsilon^{2} 
\)
and 
\(
D = {\rm det} \: (\Omega - \calH) 
\). 
We find that $C$ is strictly positive for non-zero $\Delta$. 
We also find that $D^{-1}$ is strictly negative when $\Omega$ is the
Matsubara frequency ($\Omega = i\omega_{n}$). 
The latter statement can be shown that the spectrum of the BdG
Hamiltonian is constructed by pairs of the positive and the negative
eigenvalues $(E_{\veck},-E_{\veck})$, owing to its particle-hole
symmetry. 
Therefore, when $h \to 0$, the anomalous self-energy never vanishes. 
This corresponds to the Anderson's theorem. 
In contrast, when $h$ increases, $\Sigma^{\rm A}_{\rm Born}$ can be so
small that the robustness of $T_{\rm c}$ dies out. 
Hence, the Anderson's theorem is violated, when the magnetic field is
large. 

Next, we focus on the TKNN invariant. 
The quantity is related to the number of gapless edge
modes.~\cite{Sato;Fujimoto:2010} 
The zero TKNN invariant means no gapless mode, for example. 
When the TKNN invariant is $\pm 2$, we have two modes. 
Under the open boundary condition along the $x$-direction, one mode has
a gapless behavior at $k_{y}=0$, while the other is zero at
$k_{y}=\pi$.,\cite{Sato;Fujimoto:2010} 
Thus, the non-zero TKNN invariant means the presence of low-energy edge
modes, even if $s$-wave pairing potential induces the superconductivity.

Although the $T_{\rm c}$ reduction by impurity scattering is 
a phenomena in the bulk system, we find an intriguing 
coincidence in Fig.~\ref{fig:tkkn}; the NTW-TKNN formula differs from a 
constant value, for the magnetic field generating the gapless 
edge modes on the surfaces in the clean limit. 
The non-constant behaviors of NTW-TKNN formula (\ref{eq:tkkn}) 
with respect to $h$ imply the disappearance of the robustness 
against non-magnetic impurities (See Sec.4.2). 
Therefore, we suggest that the occurrence of the gapless 
edge modes on the surfaces be relevant to the low-energy 
excitations around impurities, leading to pair-breaking effects. 
We will test this conjecture elsewhere \cite{Nagai2014}.

Before closing this section, we refer to two topics related to the
present study. 
First, we focus on an analogy with vortex physics of a chiral
$p$-wave superconductor. 
The chiral $p$-wave superconductivity has two kinds of the vortices, one of which has the vorticity parallel to the internal angular momentum of the Cooper pair, while the other of which has the anti-parallel vorticity. 
We note that the effect of non-magnetic impurities is suppressed inside a vortex core, only in the anti-parallel case \cite{Kurosawa}. 
This behavior originates from the fact that the total topological number, i.e., the summation of 
the vorticity and the chirality is zero. 
A similar effect is expected in the present topological superconductor, when the TKNN invariant is 1; 
the impurity effects inside a vortex core may depend on the direction of 
a vortex (i.e., the total topological number composed of the TKNN invariant 
and the vorticity). 
Second, a two-dimensional topological superconductor can be realized in 
superlattice structures made of $\mbox{CeCoIn}_{5}$ and
$\mbox{YbCoIn}_{5}$.~\cite{Mizukami} 
Mizukami et al.\cite{Mizukami} show that a number of the superconducting
layers can be controlled and the Rashba spin-orbit coupling can be
induced in a multi-layer structure. 
A theoretical prediction~\cite{Yoshida} suggests the
emergence of a topological spin-singlet state. 
The application of our approach to this system is an intriguing future
work.

\section{Summary}
\label{sec:summary}
In summary, we studied the non-magnetic impurity effects in a
topological $s$-wave superconductor, in terms of NTW-TKNN formula 
(\ref{eq:tkkn}). 
Our numerical calculations show that the NTW-TKNN formula is almost
constant in the presence of impurities, whenever a spectral gap opens in
the DOS. 
We examined $T_{\rm c}$ versus impurity concentrations. 
The self-consistent calculations indicate that $T_{\rm c}$ is robust
when the NTW-TKNN formula is almost constant with
respect to the Zeeman magnetic field, whereas $T_{\rm c}$ significantly
reduces when this quantity depends on the magnetic field. 
For a non-zero even TKNN invariant, a fragile $T_{\rm c}$
behavior is shown. 
Hence, we conclude that the Anderson's theorem breaks down even for
a $s$-wave superconductor and its violation is characterized well by the
value of the NTW-TKNN formula. 
The present approach suggests that intriguing effects occur via 
incorporation between a topological order and spatial inhomogeneity. 

\begin{acknowledgment}
We would like to thank A. Shitade and G. E. Volovik for helpful
comments and critical suggestions. 
The calculations were done by the supercomputing system PRIMERGY
BX900 at the Japan Atomic Energy Agency. 
This work was partially supported by JSPS KAKENHI Grant Number
 24340079. 
\end{acknowledgment}


\begin{thebibliography}{99}
\bibitem{note:selfenergy}
The second term in Eq.~(\ref{eq:S}) is the renormalization
	of $\mu$ in our $\mu$-fixed our calculations. 
For a self-consistent calculation about $\mu$, this term is not needed.
\bibitem{note:calculations}
We can confirm that our results do not change when performing the full
	self-consistent calculations.
\end{thebibliography}
\end{document}